\documentclass[preprint2]{aastex}

\newcommand{\radm}{\mbox{rad~m$^{-2}$}}
\shorttitle{SGPS: Polarized Radio Continuum Observations}
\shortauthors{Haverkorn et al.}

\begin{document}

\title{The Southern Galactic Plane Survey: Polarized Radio
  Continuum Observations and Analysis}

\author{M. Haverkorn\footnote{Jansky Fellow, National Radio Astronomy Observatory} \footnote{Current address: Astronomy Department UC-Berkeley, 601 Campbell
Hall, Berkeley CA 94720}, B. M. Gaensler\footnote{Alfred P. Sloan Fellow} \footnote{Current address: School of Physics A29, The University of Sydney, NSW 2006, Australia}
\affil{Harvard-Smithsonian Center for
    Astrophysics, 60 Garden Street, Cambridge MA 02138;
    marijke@astro.berkeley.edu, bgaensler@cfa.harvard.edu}
  N. M. McClure-Griffiths\affil{Australia Telescope National
    Facility, CSIRO, PO Box 76, Epping, NSW 1710, Australia;
    naomi.mcclure-griffiths@csiro.au} 
  J. M. Dickey\affil{Physics Department, University of Tasmania,
  Private Bag 21, Hobart TAS 7001, Australia; john.dickey@utas.edu.au}
  A. J. Green\affil{School of Physics, University of Sydney, NSW 2006, Australia; agreen@physics.usyd.edu.au}}

\begin{abstract}
The Southern Galactic Plane Survey (SGPS) is a radio survey in the 21
cm \ion{H}{1} line and in 1.4~GHz full-polarization continuum,
observed with the Australia Telescope Compact Array and the Parkes 64m
single dish telescope. The survey spans a Galactic longitude of
$253\degr < l < 358\degr$ and a latitude of $|b|<1\degr$ at a
resolution of 100 arcsec and a sensitivity below 1~mJy/beam. This
paper presents interferometer only polarized continuum survey data and
describes the data taking, analysis processes and data products. The
primary data products are the four Stokes parameters I, Q, U, and V in
25 overlapping fields of $5\degr.5$ by $2\degr$, from which polarized
intensity, polarization angle and rotation measure are calculated.  We
describe the effects of missing short spacings, and discuss the
importance of the polarized continuum data in the SGPS for studies of
fluctuations and turbulence in the ionized interstellar medium and for
studying the strength and structure of the Galactic magnetic field.
\end{abstract}


\section{Introduction}

Galactic magnetism is one of the major components in the Milky Way,
mostly in equipartition with gas and cosmic rays.
Interstellar magnetic fields are believed to profoundly influence the
ionized interstellar medium (ISM) through flux freezing and energy
dissipation, affect star formation, determine the trajectories and
acceleration of low and medium energy cosmic rays and play a major
role in the turbulent gas dynamics (see e.g. reviews by Ferri\`ere
2001, Scalo \& Elmegreen 2004, Elmegreen \& Scalo 2004).

Knowledge about the strength and structure of the Galactic magnetic
field is still sketchy. Yet, the field has received increasing
attention, not only with the objective of studying magnetic fields in
galaxies but also because the Galactic magnetized ISM forms a
polarized foreground which needs to be determined for Cosmic Microwave
Background Polarization \citep{dto03} and Epoch of Reionization
studies \citep{mh04}.

The only methods to probe Galactic magnetic fields in diffuse ionized
gas over a large range of spatial scales are by way of radio
polarization and Faraday rotation. Right and left circularly polarized
components of radio emission experience birefringence while
propagating through a magnetized and ionized medium. This causes the
polarization angle of linearly polarized emission $\phi$ to rotate as
a function of wavelength $\lambda$ as $\Delta\phi =
\mbox{RM}\lambda^2$, where the rotation measure RM is RM~$=0.81\int
n_e \mathbf{B}\cdot\mathbf{ds}$, $n_e$ is the thermal electron density
in cm$^{-3}$, $\mathbf{B}$ is the magnetic field vector in microGauss,
$\mathbf{ds}$ is the path length vector through the medium in parsecs,
and the integral is along the line of sight from the observer to the
source of polarized emission. Therefore, Faraday rotation measurements
allow estimation of the magnetic field component along the line of
sight, weighted by the electron density, and integrated over the
pathlength. Depolarization characteristics can be used to determine
the scale and amplitude of fluctuations in the medium.

The observed polarized radiation used to trace the Galactic magnetic
field can come from pulsars, polarized extragalactic sources or
diffuse Galactic synchrotron emission (including supernova
remnants). All of these sources have their own advantages and
disadvantages. Pulsars are unique because model-dependant distance
estimates allow constraining of the path length, and because a
dispersion measure can be calculated, which in combination with RM
yields a direct measure of the magnetic field averaged over the path
length. However, they are scarce and distributed mainly in the
Galactic plane. Unresolved extragalactic sources, on the other hand,
are distributed all over the sky. But they have an intrinsic RM
contribution, and currently published datasets yield an average of one
source~deg$^{-2}$ in the Galactic plane (Brown et al.\ 2003, Brown et
al., in prep), and only 0.02-0.03 source~deg$^{-2}$ in the rest of the
sky (e.g.\ Simard-Normandin et al.\ 1981, Broten et al.\ 1988). Only
diffuse synchrotron emission provides a pervasive background of
polarized radiation which can be used to form RM maps of large fields
in the sky with high resolution \citep{hkb03a, hkb03b, ulg03,
r04}. Diffuse synchrotron emission does suffer from depolarization,
which decreases the possible measurements of RM, but which in itself
can yield information about the fluctuations in the magneto-ionized
interstellar medium \citep{gdm01}.

Recently the whole sky has been mapped in absolutely calibrated
polarized continuum at 1.4~GHz (see Wolleben et al.\ 2005 for the
Northern sky, and Testori et al.\ 2004 in the South) at a resolution
of about half a degree. More than half of the Galactic plane is being
surveyed at the much higher resolution of an arcmin, in two separate
polarization surveys.  The Canadian Galactic Plane Survey (CGPS,
Taylor et al.\ 2003) covers the Northern Sky at Galactic longitudes
$74.2\degr <l< 147.3\degr$ and latitudes $-3.6\degr<b<+5.6\degr$. The
Southern Galactic Plane Survey (SGPS, McClure-Griffiths et al.\ 2005;
hereafter Paper I) extends from $253\degr < l < 358\degr$ (Phase~I)
and $5\degr <l<20\degr$ (Phase~II) at longitudes $|b|<1\degr$. In both
these projects full-polarization continuum and \ion{H}{1} were
observed simultaneously. Although for the \ion{H}{1} part of the SGPS
data from the Australia Telescope Compact Array (ATCA) interferometer
and Parkes 64-m single dish have been combined, the polarization
measurements discussed here are derived from ATCA observations only.

The \ion{H}{1} data in the SGPS have been discussed in detail in
Paper~I, whereas a test region covering the region $325\degr.5 < l <
333\degr.5$ and $-0\degr.5 < b < +3\degr.5$ is described in
\citet{gdm01} and \citet{mgd01}.  This paper describes the radio
continuum data of the SGPS Phase~I, i.e.\ at Galactic longitudes
$253\degr < l < 358\degr$ (hereafter refered to as SGPS); the
continuum data observed in the region $5\degr<l<20\degr$ will be
discussed elsewhere. In Sect.~\ref{s:data} we briefly summarize some
details about the telescopes, data taking, and
calibration. Furthermore, this Section describes the polarization
calibration and the effect of missing short spacings.  In
Section~\ref{s:res} we present the data, and Sect.~\ref{s:science}
discusses some of the science done with the polarized SGPS data.

\section{The Survey}
\label{s:data}

\begin{table*}[t]
\begin{center}
\begin{tabular}{l|c|c|c|c|c|c|c}
\hline
\bf Source    & \bf RA        & \bf dec & \bf l   & \bf b   &
\bf I$_{1.4}$ & \bf no.\ pointings & \bf time \\
              & (h:m:s)   & (\degr:m:s) & (\degr) & (\degr) &
(Jy)          &               & (hr)       \\
              & (J2000.0) & (J2000.0)   &         &         &
              &               &            \\
\hline\hline
PKS 0857-43    & 08:59:28 & -43:45:44 & 265.15 & +1.45 &  26 &  7 & 1.7\\
CTB 31         & 08:59:07 & -47:31:24 & 267.95 & -1.06 & 100 &  7 & 1.7\\
PKS 0922-51    & 09:24:26 & -51:59:34 & 274.01 & -1.15 &  12 &  6 & 1.7\\
GAL 282.0-01.2 & 10:06:38 & -57:12:11 & 282.02 & -1.18 &  11 &  7 & 1.7\\
Gum 29         & 10:24:15 & -57:46:58 & 284.31 & -0.33 &  73 &  7 & 1.7\\
GAL 285.3-00.0 & 10:31:29 & -58:02:08 & 285.26 & -0.05 &  12 &  7 & 1.5\\
NGC 3372       & 10:44:19 & -59:37:34 & 287.61 & -0.85 & 158 &  7 & 1.5\\
NGC 3581       & 11:11:57 & -61:18:49 & 291.28 & -0.71 &  37 & \raisebox{-1ex}{$\}$10} & \raisebox{-1ex}{$\}$2.1}\\
GRS 291.6-00.5 & 11:15:10 & -61:16:45 & 291.63 & -0.54 &  22 &    & \\
GAL 298.2-00.3 & 12:10:03 & -62:50:00 & 298.23 & -0.34 &  16 & \raisebox{-1ex}{$\}$13} & \raisebox{-1ex}{$\}$3.0}\\
GAL 298.9-00.4 & 12:15:26 & -63:01:28 & 298.86 & -0.44 &  15 &    & \\
GAL 305.4+00.2 & 13:12:33 & -62:34:43 & 305.36 & +0.19 &  18 & 10 & 2.4\\
GRS 307.1+01.2 & 13:26:19 & -61:23:01 & 307.10 & +1.21 &  10 &  7 & 1.7\\
GAL 309.6+00.1 & 13:46:49 & -60:24:29 & 309.72 & +1.73 & 119 &  7 & 1.7\\
GAL 316.8-00.1 & 14:45:19 & -59:49:32 & 316.80 & -0.06 &  25 &  7 & 1.7\\
GAL 327.3-00.5 & 15:52:35 & -54:38:00 & 327.30 & -0.56 &  31 &  7 & 1.5\\
GRS 326.7+00.6 & 15:44:48 & -54:06:42 & 326.66 & +0.57 &  21 &  7 & 1.5\\
GAL 331.5-00.0 & 16:12:03 & -51:26:55 & 331.52 & -0.07 &  25 &  7 & 1.5\\
GRS 333.6-00.2 &16:22:10 & -50:06:07 & 333.60 & -0.22 &  38 &  7 & 1.4\\
\end{tabular}
\caption{Additional observations to increase SGPS dynamic range around
         bright sources. From left to right, the columns denote the
         source's name, its coordinates in RA and dec, its Galactic
         longitude and latitude, its integrated flux density at
         1.4~GHz, the number of pointings and the total amount of time
         spent observing this source.\label{table1}}
\end{center}
\end{table*}

\subsection{Observations}

The ATCA is a radio interferometer near Narrabri, NSW, Australia,
consisting of six 22m-diameter dishes. Five of these are located on a
3km long east-west track, or a 200m north-south track, while the sixth
element is fixed 3km west of the east-west track, so that baselines
from 31m to 6km can be achieved. The SGPS was observed only on
east-west baselines, between 1998 December and 2001 June. As the main
objective of the SGPS is measuring diffuse emission, we exploited the
compact configurations of the instrument. The 6km antenna was excluded
and five different compact configurations were used to ensure
continuous uv-coverage at baselines from approximately 31m to 3km,
which results in a resolution of $\sim80$~arcsec (smoothed to
100~arcsec for all fields) and sensitivity up to scales of about
30~arcmin.

The ATCA data signal is processed through two intermediate frequency
(IF) channels, allowing observations in the \ion{H}{1} band and a
continuum band simultaneously. Data in the continuum band were
averaged into 12 frequency channels of each 8~MHz wide, centered on
1336~MHz to 1432~MHz, where the 1408~MHz band was flagged due to
internal interference. Two feeds $X$ and $Y$ detect orthogonal linear
polarizations, which are correlated into signals in four channels
$XX$, $YY$, $XY$ and $YX$ for each antenna pair. These were translated
into Stokes parameters I, Q, U, and V \citep{hb96}.

The SGPS data are divided into fields of 5\degr.5 in Galactic
longitude and 2\degr\ in Galactic latitude, centered on $b=0\degr$,
and slightly overlapping in longitude. The central longitudes of the
fields run from 255\degr\ to 355\degr\ in steps of $5\degr$. Each
field consists of 105~pointings in a hexagonal pattern with Nyquist
spacing of 19\arcmin\ between pointing centers. This results in a
constant gain across fields except at the edges of the survey, and
decreases the contribution of instrumental polarization to about
0.1\%. Each pointing was observed for 30~sec before continuing to the
next pointing, circling through a field for 12 hours. Multiple 12-hour
runs assured regular uv-coverage and a total observing time per
pointing of at least 20~min. See Paper~I for observational details
about the SGPS data set such as exact locations and sizes of the
fields and pointings, uv-coverage etc.

Calibration, deconvolution and flagging eliminated most artifacts in
the field. However, grating rings with a radius of $0.8\degr$ can
still be seen around bright and extended sources, mostly in total
intensity but occasionally also in polarization, sometimes accompanied
by spokes. This is due to the regular spacing of about 15m in the
telescope configurations of the east-west array, combined with the
relatively low dynamic range of the SGPS of approximately 150:1 (see
Paper~I). The bright sources and/or their artifacts can induce
instrumental polarization if strong enough. To mitigate these
artifacts, we took additional observations of strong and extended
sources in or near the field of observation which increased the
dynamic range in these pointings. These sources were observed in small
mosaics with the same spacing, pattern and approximate observing time
as the pointings in the SGPS, as detailed in Table~\ref{table1}. The
noise in the final Q, U images is $\sim0.3$~mJy/beam at low Galactic
longitudes, and increases to $\sim0.6$~mJy/beam towards the Galactic
Center. The average error in polarization angle on the total band is
3\degr, the average error in angle in one 8~HMz channel is 10\degr.

The data were reduced using the software package MIRIAD
\citep{sk03}. Flux and bandpass variations between antennas were
calibrated using the primary calibrator PKS~B1934-638, assuming a flux
of 14.9~Jy at 1.4~GHz (Reynolds 1994). The primary calibrator was
observed for $\sim10$~min at the beginning and end of each observing
session. For each program field, a secondary calibrator close to the
field was observed. Atmospheric gain and delay variations could be
monitored and calibrated using the secondary calibrators. For some
fields the secondary calibrator was observed regularly during the
night so that sufficient hour angle coverage could be obtained and the
calibrator's intrinsic polarization could be separated from the
instrumental polarization leakage. For the remaining fields the hour
angle range of the secondary calibrator was too small to obtain
polarization calibration so that 1934-628 was used instead, assuming
it is unpolarized at 1.4~GHz.  These leakage and gain parameters per
antenna, combined with the bandpass calibrations from the primary
calibrator, were propagated to the program fields. Additional flagging
per pointing and/or per Stokes parameter was necessary to eliminate
bad data due to interference. For details about the calibrators and
non-polarization calibration, see Paper~I.

The pointings were linearly combined for each 11~deg$^2$ field and
imaged jointly using a standard grid-and-FFT scheme. Superuniform
weighting was used for maximum reduction of grating rings, at the
expense of signal-to-noise. The maps were deconvolved jointly, which
allows some of the missing large-scale structure to be retrieved
(Sault et al.\ 1996). Using a maximum entropy algorithm (Narayan \&
Nityananda 1984), the Stokes parameters can be deconvolved jointly
(routine PMOSMEM in MIRIAD) or separately (MOSMEM). As has been noted
by \citet{sbd99} and confirmed with the SGPS data, for diffuse
polarization mostly uncorrelated to total intensity joint deconvolution
is mildly inferior to separate deconvolution.  A deconvolution
algorithm for mosaics based on Steer CLEANing (MOSSDI, Steer et
al.\ 1984) did not give significantly better results, therefore
maximum entropy deconvolution for all Stokes parameters separately was
used. The images were restored with a Gaussian beam of 100~arcsec
resolution. 

\subsection{Missing short spacings}

An interferometer is insensitive to structure on large scales due to
a finite minimum spacing between dishes. Therefore, the SGPS does not
contain structure on scales larger than about 30~arcmin.

Missing large-scale structure in Stokes Q and U can destroy or create
small scale structure in polarized intensity \citep{hkb04a},
i.e. enhancements in polarization on small scales can become
depressions if zero spacing data is added, or vice versa.
Furthermore, due to the non-linear dependence of polarization angle on
Stokes Q and U, the linear relation between polarization angle and
wavelength squared caused by Faraday rotation may be destroyed,
although this is not always the case. Specifically, if the large scale
emission is much stronger than the small scale emission, the computed
RM will deviate from the true value.  Therefore, addition of
single-dish structure (Stanimirovi\'c 2002) to polarized emission from
interferometers is imperative (Uyan\i ker et al. 1998).

Although we plan to eventually include single-dish polarization data
from the Parkes 64m telescope in the SGPS, the polarization data
discussed in this paper do not contain large-scale structure. This
introduces important caveats associated with the use of these data, as
we now discuss.

Spatial RM variations on scales smaller than the sensitivity scale of
the interferometer ($< 0.5\degr$) cause a spread in polarization angle
within the field of view. If the RM variations are large enough to
cause a variation in polarization angle which is higher than about
$\pi/2$~radians, this will effectively randomize Q and U within the
field of view, destroying any large-sale structure in Q and U
(Haverkorn et al.\ 2004a). Therefore, an interferometer can detect all
of the Q and U structure if RM variations within its field of view are
large enough. In the case of the ATCA, RM variations of about
40~\radm\ or larger on scales smaller than the field of view cause a
sufficiently large spread in polarization angle that Q and U are
effectively randomized over the field of view, so that no large-scale
structure in Q and U remains (even though large-scale structure in RM
is present). The observed value of RM variations is similar to this
critical value.

Radiation from unresolved extragalactic point sources and pulsars does
not have a large-scale component. Any contribution from diffuse
foreground emission to the polarized emission in the direction of the
point source can be estimated and subtracted using the source's
immediate surroundings \citep{btj03}. Therefore, missing large-scale
structure will not influence polarized point source measurements.

Care has to be taken in interpretation of the diffuse emission. For
structures that emit on small scales detectable by the interferometer
(e.g. the Vela supernova remnant) it can be assumed that the RM
measured from the emitting structures is accurate. However, for
structures produced by pure Faraday rotation it is not certain whether
the computed RM is equal to the real RM. For most pixels no linear fit
between $\phi$ and $\lambda^2$ could be found, indicating
depolarization and/or missing large-scale structure. Only for a small
fraction of the SGPS data (typically a few percent) the calculated
linear fit for $\phi$ against $\lambda^2$ yields a reduced $\chi^2<2$
and a signal-to-noise~$>5$. Where this structure does not coincide
with small-scale emission, RMs may deviate from the true values.

However, even even where large-scale polarized structure is missing
the SGPS can be used as a detection survey. The polarization data
contain a variety of 'objects' that are invisible in total intensity
and unrelated to emission at any other wavelength. These polarized
objects can be followed up with single-dish observations to measure
their large-scale emission, which will allow detailed study.

\section{Results}
\label{s:res}

The resulting primary data products from the ATCA are maps of Stokes
I, Q, U and V, from which debiased polarized intensity $P$
and polarization angle $\phi$ are constructed as
\begin{eqnarray}
P    &=& \sqrt{Q^2+U^2-\sigma^2} \\
\phi &=& 0.5 \arctan \frac{U}{Q} 
\end{eqnarray}
\citep{wk74} where $\sigma$ is the rms noise in the Q and U maps,
computed as the standard deviation of Stokes~$V$.

Total intensity Stokes~I, Stokes~Q, Stokes~U and polarized intensity P
maps for the complete SGPS are shown in Fig.~\ref{f:ipiqu}. For the
first field shown, Stokes V is added for illustration purposes. Some
polarization coinciding with strong sources in total intensity is
artifical and caused by instrumental polarization, down to a level of
0.1\%.  The strongest artifacts are marked with circles in
Fig.~\ref{f:ipiqu}. The remaining extended structure is believed to be
real, and the lack of correlation with total intensity indicates that
the fluctuations in polarized emission are due to depolarization and
Faraday rotation effects.

In a large part of the maps there is small-scale total intensity but
no polarized emission, indicating that the synchrotron radiation is
completely depolarized due to the abundant small-scale structure in
magnetic field and/or electron density in the inner Galactic
plane. Depolarization due to structure in electron density is
evidenced by an anticorrelation between polarization and H$\alpha$
emission frequently observed in the SGPS, see Fig.~\ref{f:shassa}.

\begin{figure*}
  \includegraphics[scale=1]{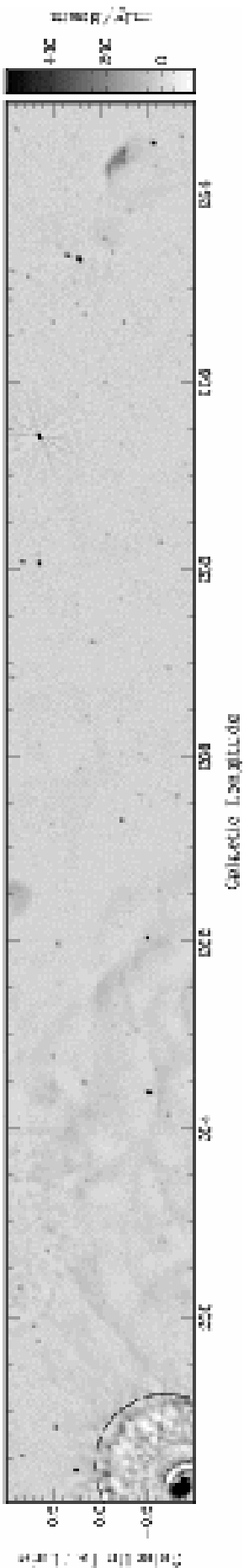}
  \includegraphics[scale=1]{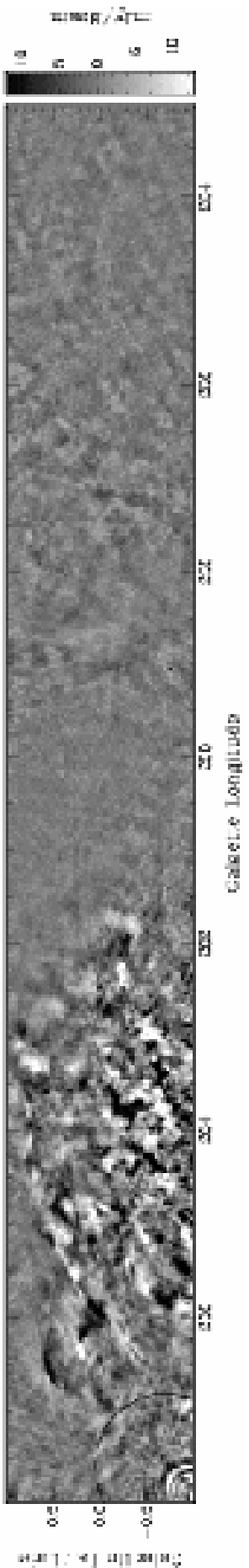}
  \includegraphics[scale=1]{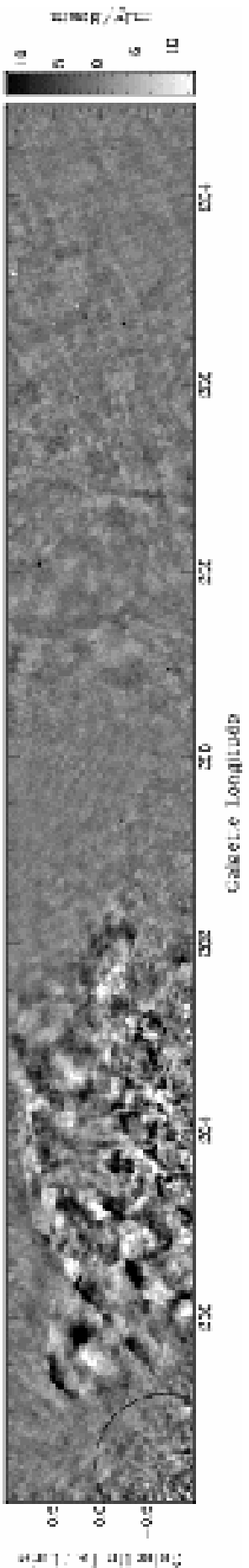}
  \includegraphics[scale=1]{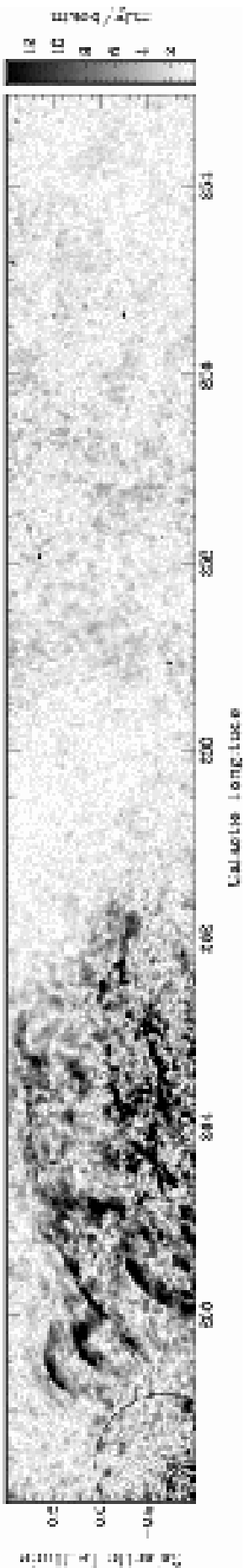}
  \includegraphics[scale=1]{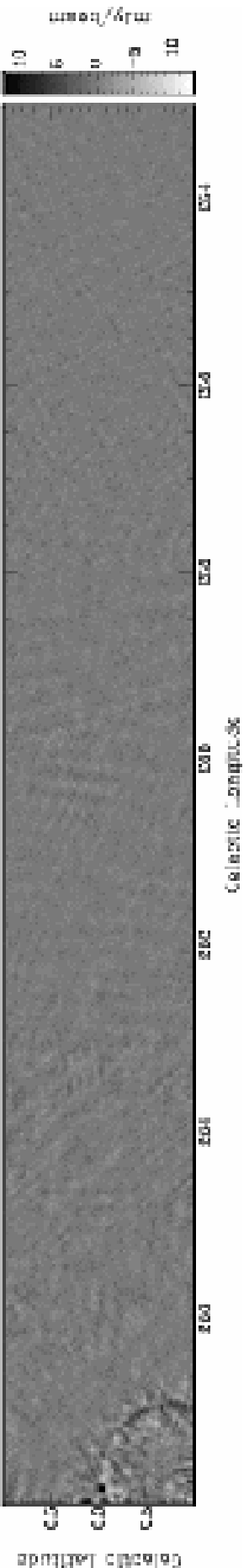}
  \caption{The Southern Galactic Plane Survey total intensity Stokes
           I, Stokes Q, Stokes U and polarized intensity P from top to
           bottom. On the first page only a Stokes V map has been
           added to the bottom. Overlaid circles denote artifacts in
           the data due to the regular 15m spacing of the ATCA, and no
           large-scale structure ($>0.5\degr$) is included. The top
           half of a weak shell centered at $l\approx 264\degr$ is the
           Vela supernova remnant.}
  \label{f:ipiqu}
\end{figure*}

\setcounter{figure}{0}

\begin{figure*}[p]
  \includegraphics[scale=1]{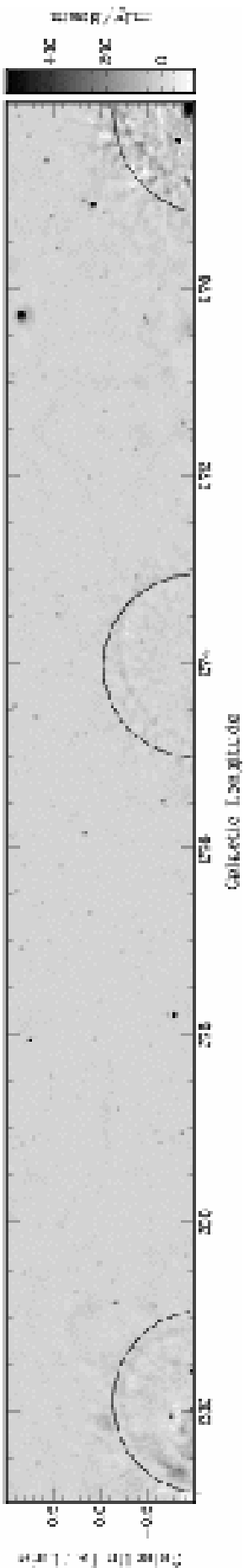}
  \includegraphics[scale=1]{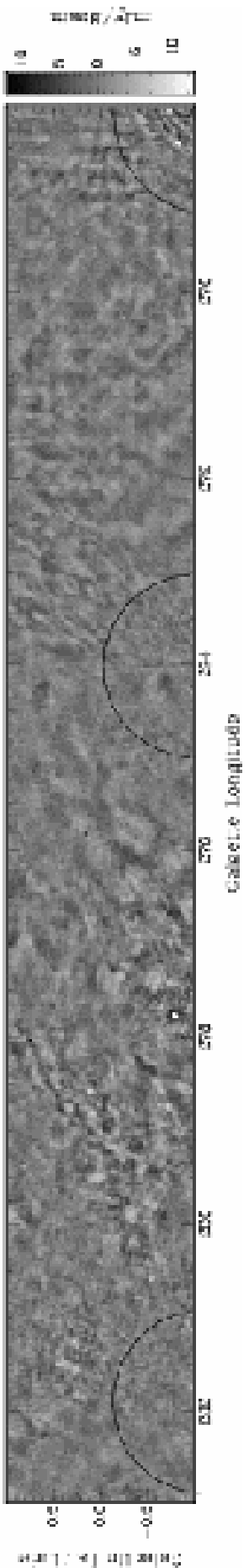}
  \includegraphics[scale=1]{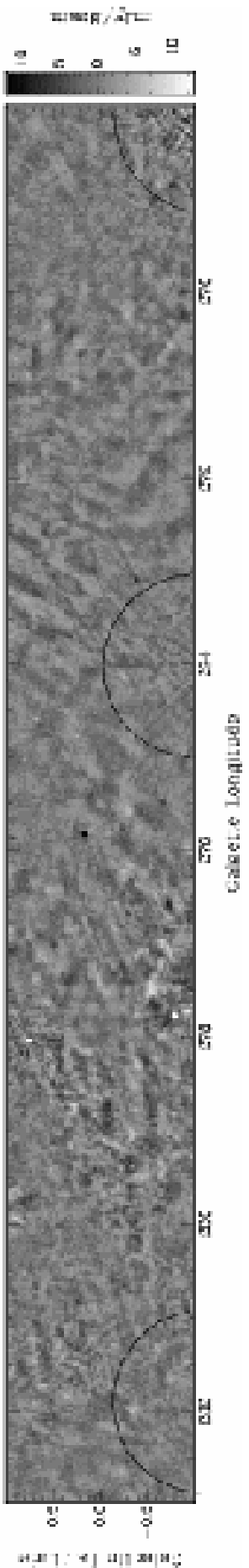}
  \includegraphics[scale=1]{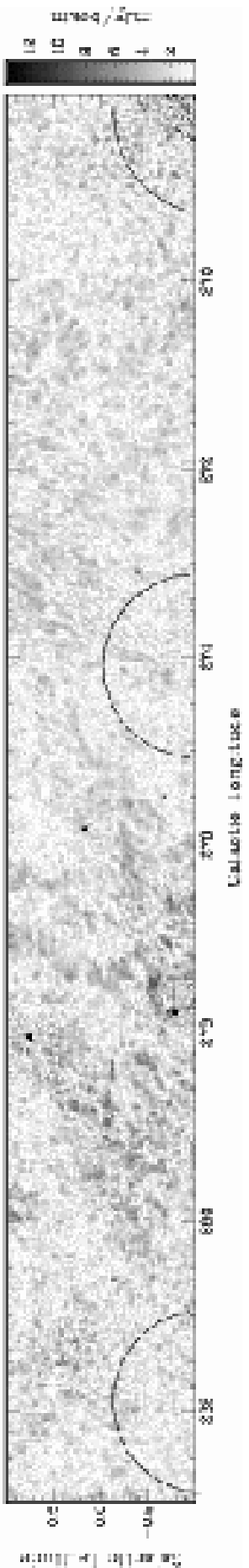}\\
  \caption{--- continued.}
\end{figure*}

\setcounter{figure}{0}

\begin{figure*}[p]
  \includegraphics[scale=1]{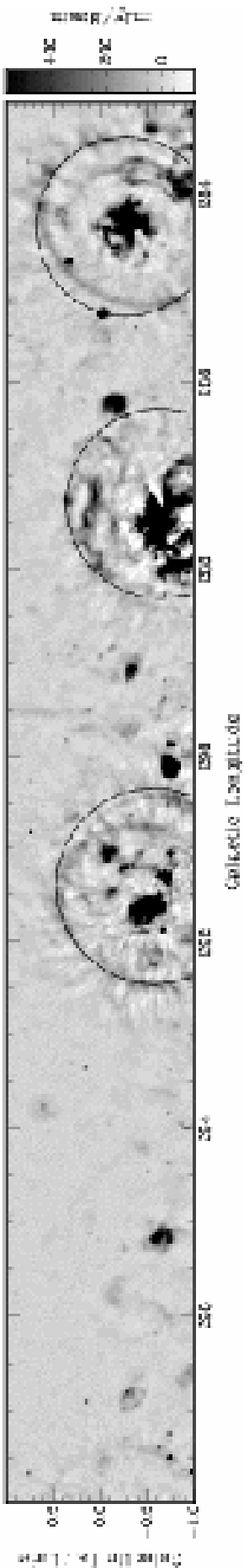}
  \includegraphics[scale=1]{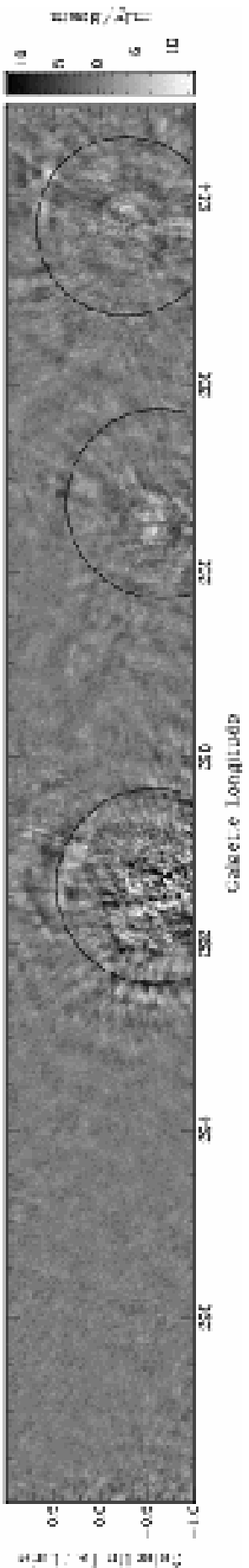}
  \includegraphics[scale=1]{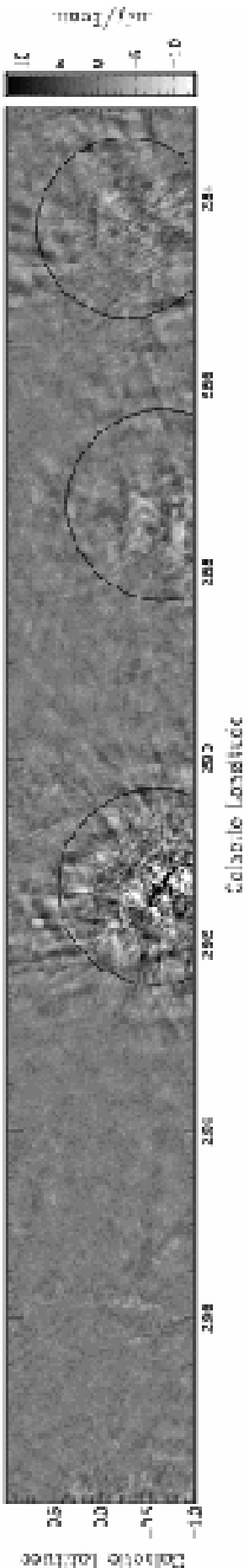}
  \includegraphics[scale=1]{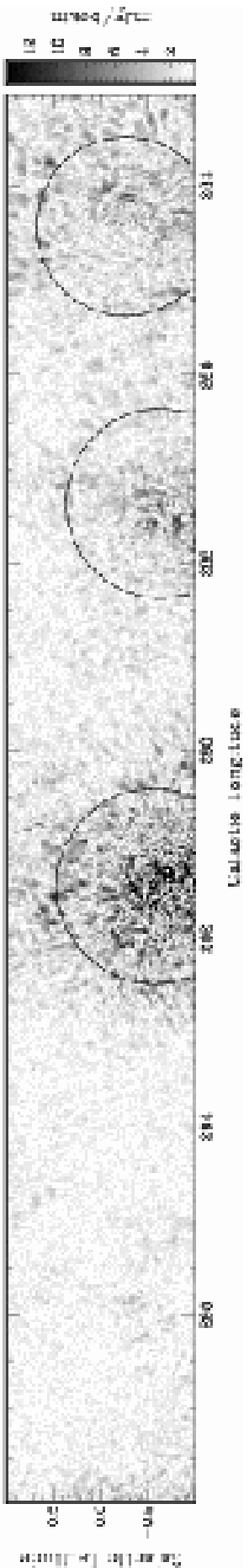}\\
  \caption{--- continued.}
\end{figure*}

\setcounter{figure}{0}

\begin{figure*}[p]
  \includegraphics[scale=1]{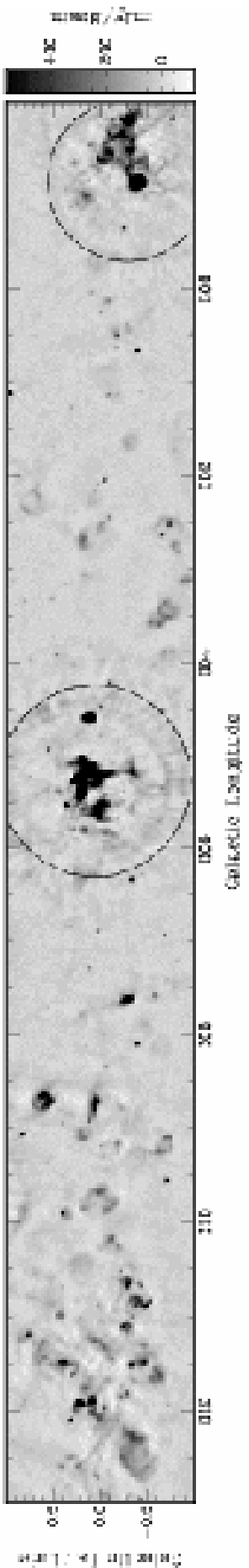}
  \includegraphics[scale=1]{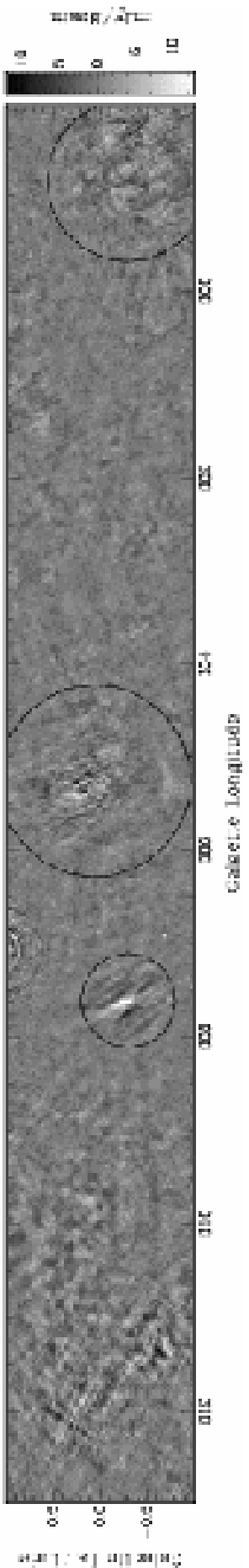}
  \includegraphics[scale=1]{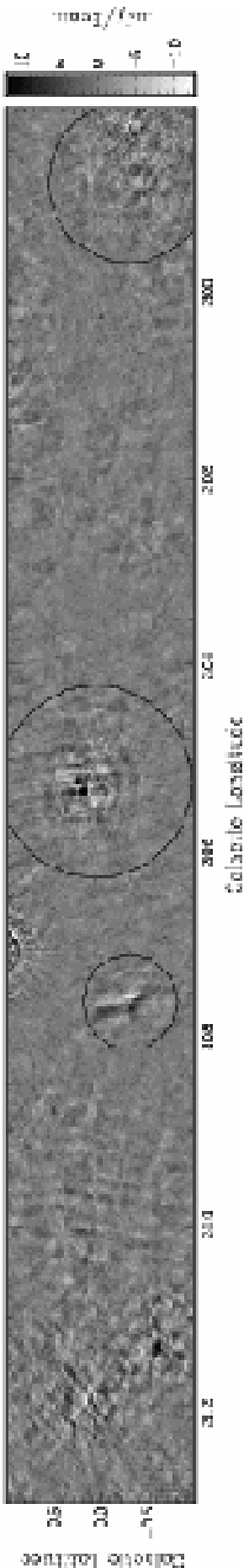}
  \includegraphics[scale=1]{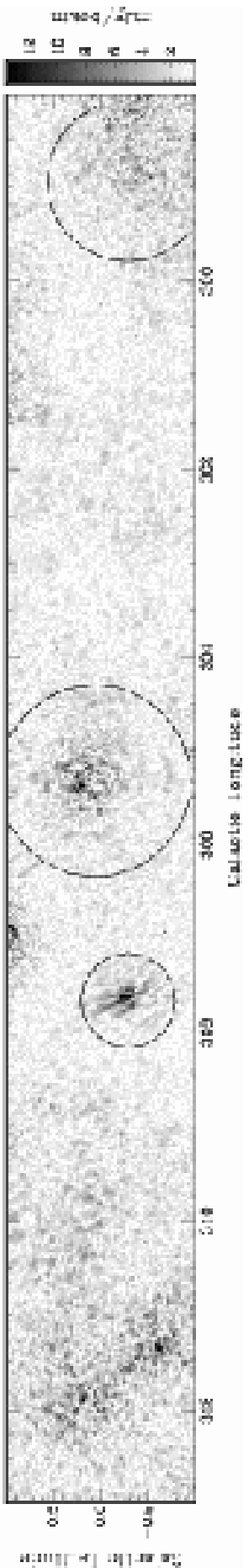}\\
  \caption{--- continued.}
\end{figure*}

\setcounter{figure}{0}

\begin{figure*}[p]
  \includegraphics[scale=1]{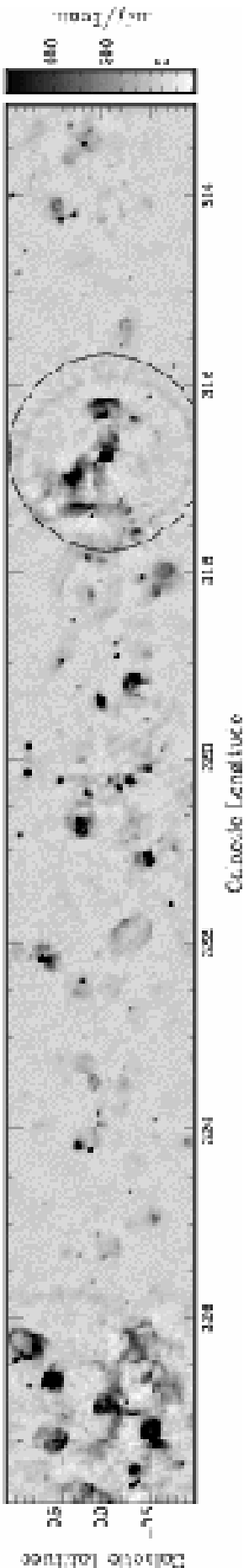}
  \includegraphics[scale=1]{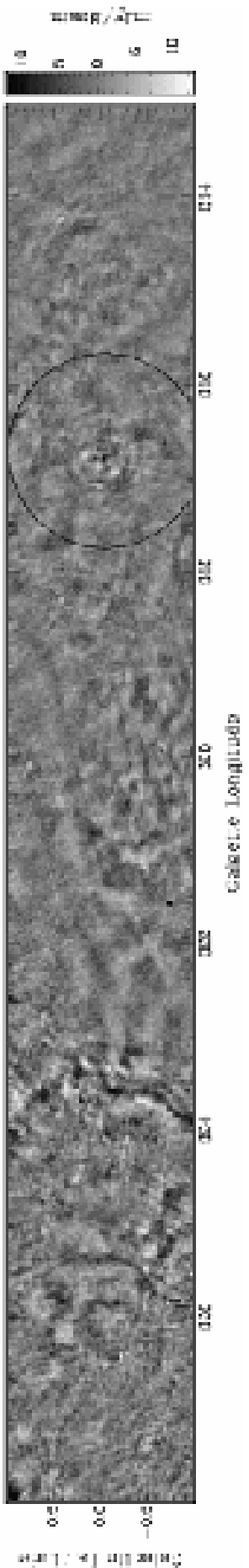}
  \includegraphics[scale=1]{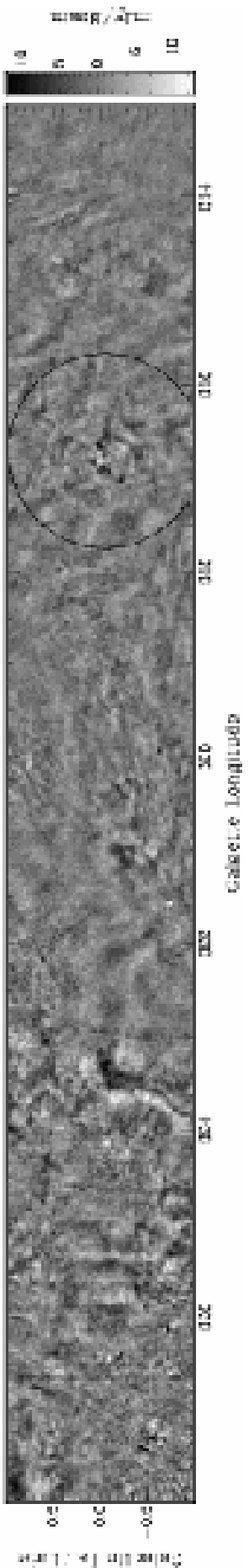}
  \includegraphics[scale=1]{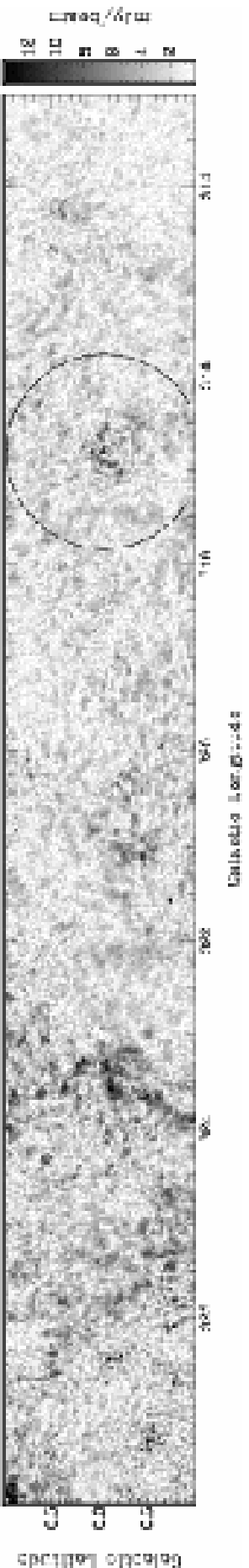}\\
  \caption{--- continued.}
\end{figure*}

\setcounter{figure}{0}

\begin{figure*}[p]
  \includegraphics[scale=1]{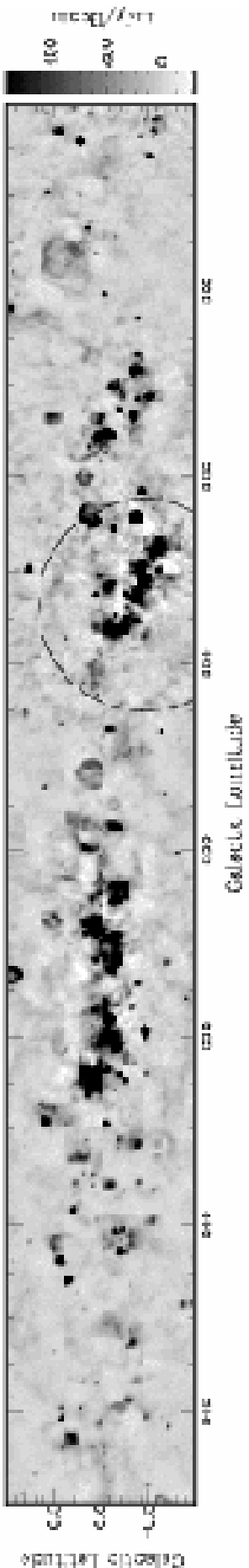}
  \includegraphics[scale=1]{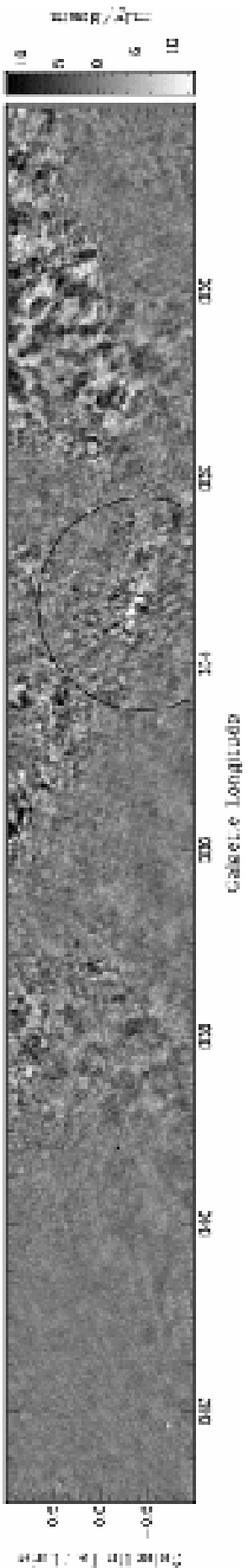}
  \includegraphics[scale=1]{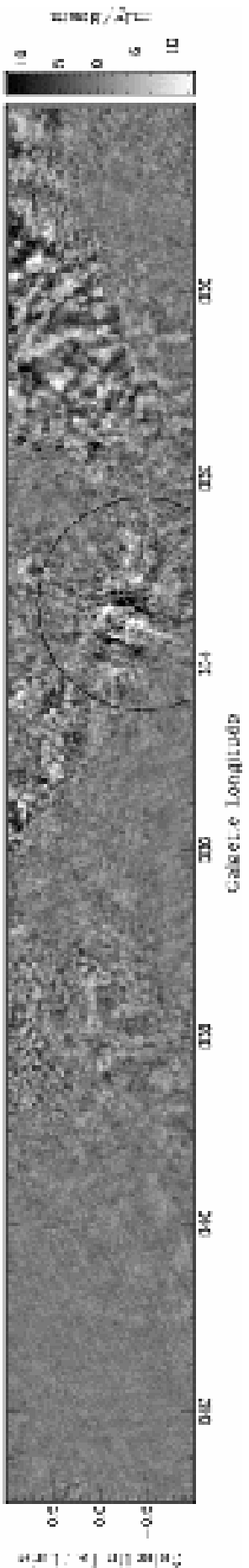}
  \includegraphics[scale=1]{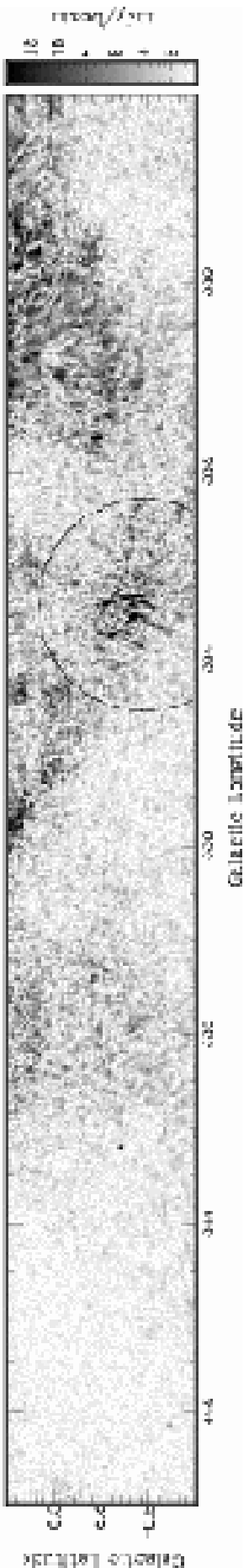}\\
  \caption{--- continued.}
\end{figure*}

\setcounter{figure}{0}

\begin{figure*}[p]
  \includegraphics[scale=1]{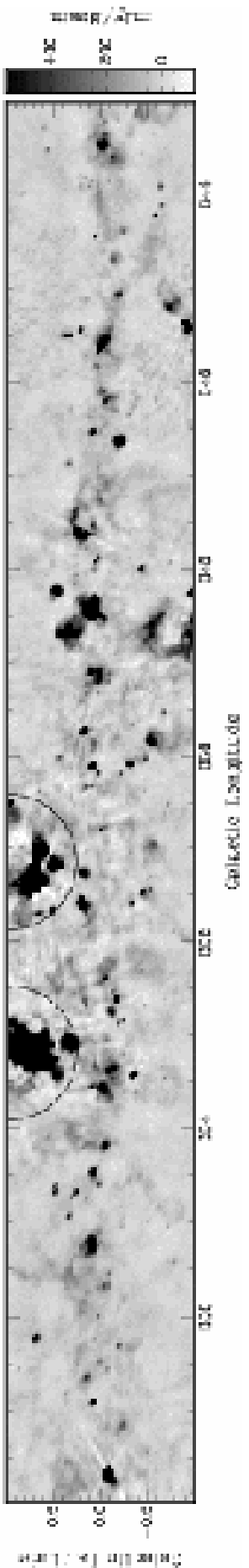}
  \includegraphics[scale=1]{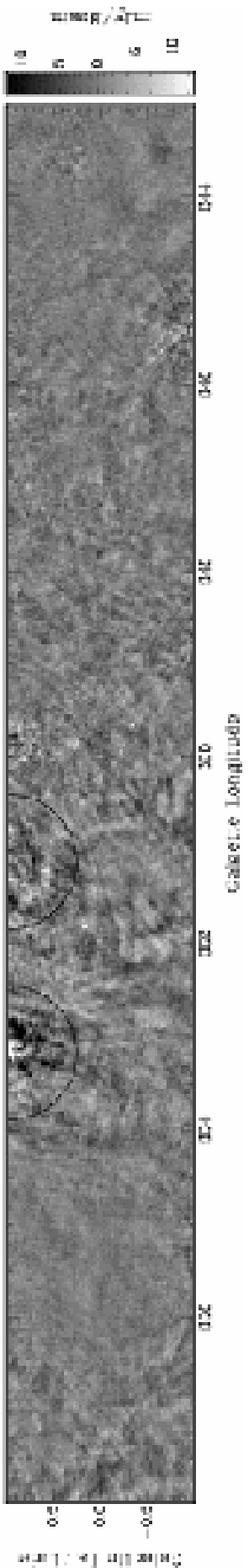}
  \includegraphics[scale=1]{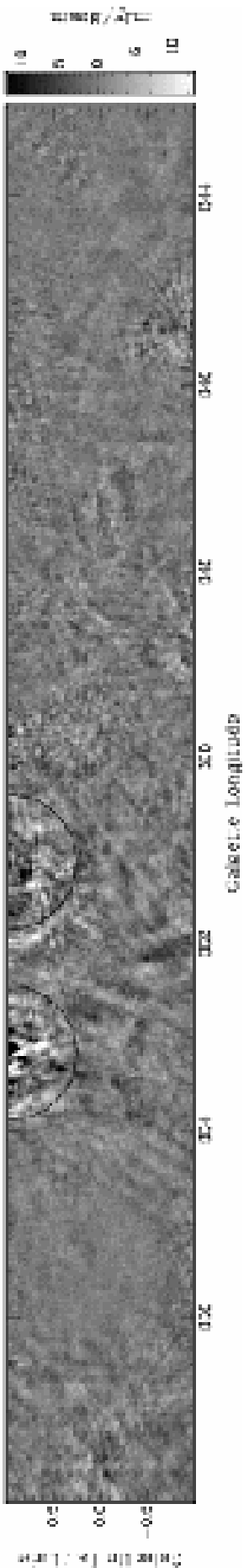}
  \includegraphics[scale=1]{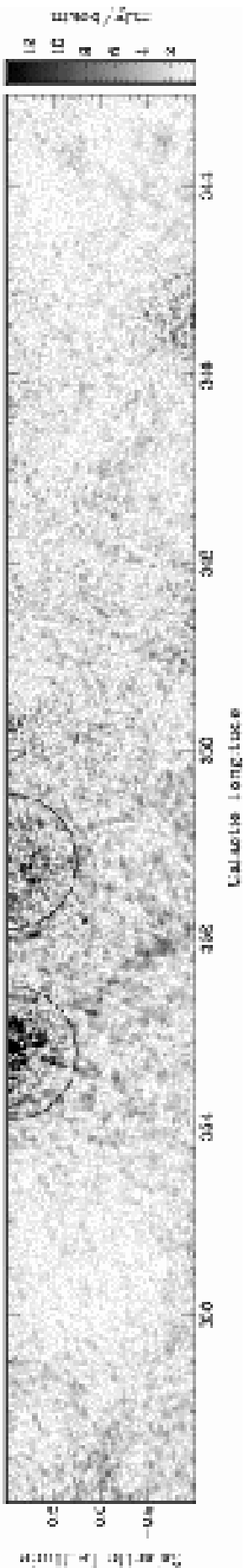}\\
  \caption{--- continued.}
\end{figure*}

\begin{figure*}[t]
  \epsscale{1.9}
  \plotone{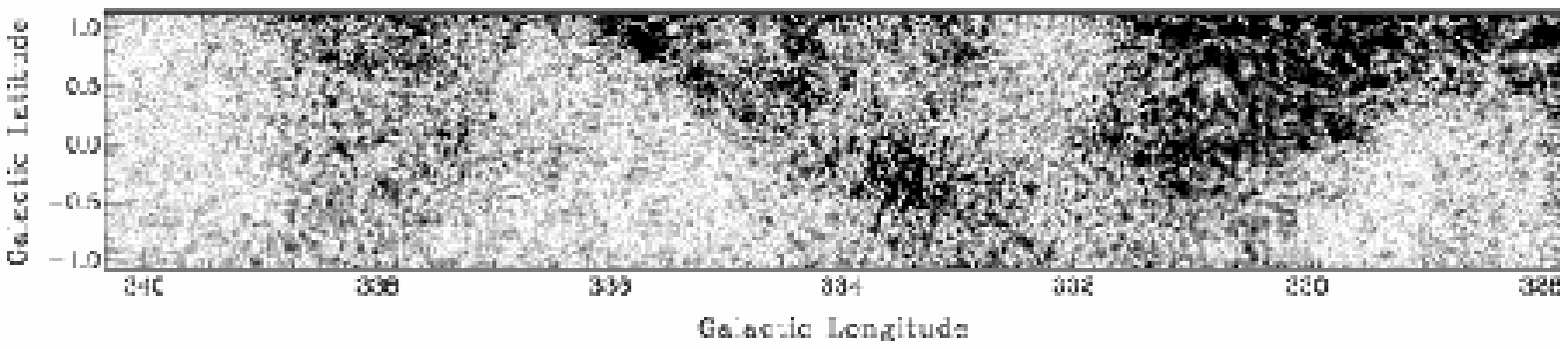}\\
  \plotone{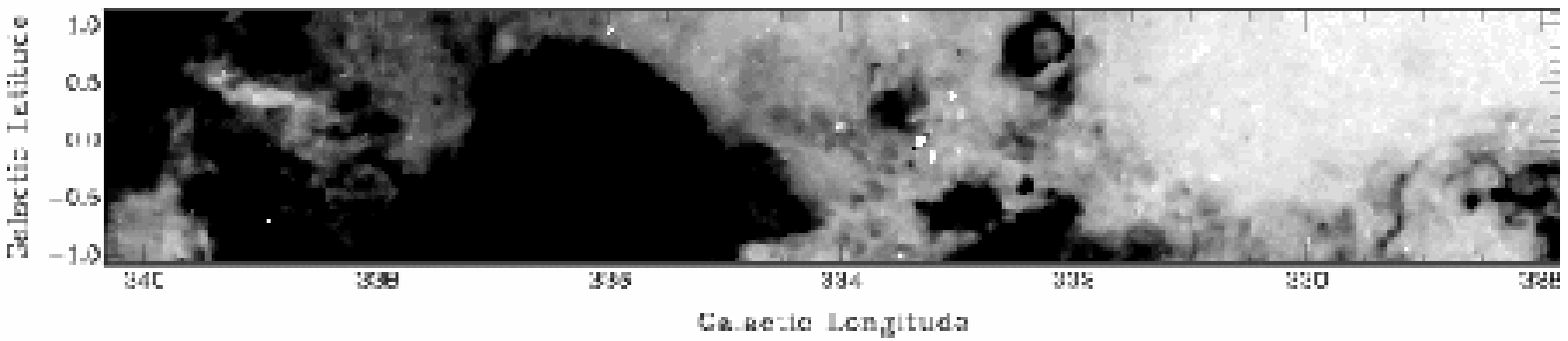} 
  \caption{Anticorrelation between SGPS polarized intensity (top) and
    H$\alpha$ emission from the SHASSA survey (bottom). The circle
    denotes artifacts in the polarization data. High H$\alpha$
    intensity is saturated to show the weak emission better.}
  \label{f:shassa}
\end{figure*}

\begin{figure*}
  \epsscale{2}
  \plotone{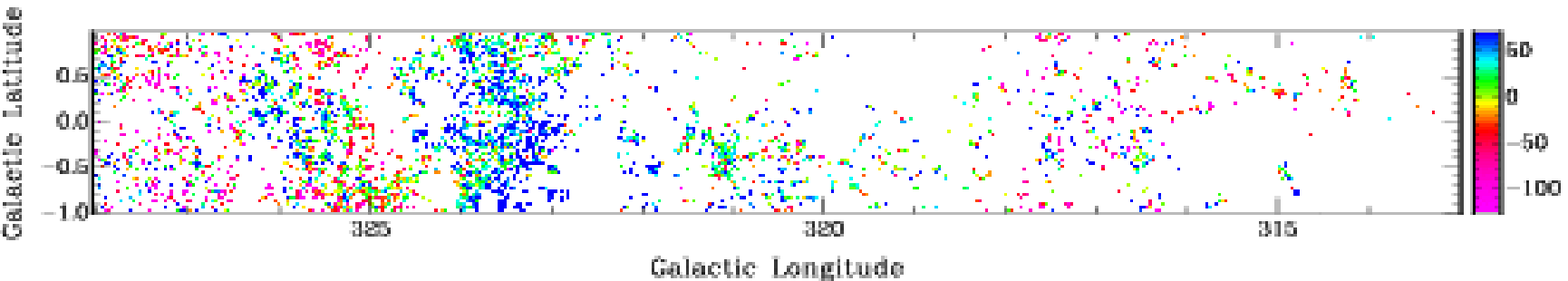}
  \caption{Example of the SGPS RM data, obtained with the PACERMAN
           algorithm \citep{dve05,vde05}.}
  \label{f:rm}
\end{figure*}

From the polarization angle maps at 12 frequencies, rotation measure
RM can be computed as $\phi = \mbox{RM} \lambda^2$, an example of
which is shown in Fig.~\ref{f:rm}. This can be done using all 12
frequency channels separately, or averaging them into 3 wider
frequency bands, for which the average error in the polarization
angle is $\sim3^{\circ}$, which give similar results. RM has been
calculated using the algorithm PACERMAN \citep{dve05,vde05}, which
computes RM values not only taking into account the $\pi$~rad
ambiguity of polarization angle, but also consistency with the
surrounding pixels. The results obtained with the PACERMAN algorithm
were checked against maps obtained by simple pixel-by-pixel linear
fits of $\phi$ against $\lambda^2$. The results are very similar,
although the PACERMAN maps look smoother than the pixel-by-pixel maps,
and have slightly more pixels for which RM could be calculated.

\subsection{Available data products}

Maps of Stokes I, Q and U are available from the SGPS Data
Server\footnote{http://www.atnf.csiro.au/research/HI/sgps/queryForm.html},
consisting of all 12 frequency channels averaged into one map. Data in
individual frequency channels and rotation measure information are
available upon request.

\section{Science with the SGPS}
\label{s:science}

From the SGPS test region, \citet{gdm01} showed that the brightest
polarized region originated in the Crux spiral arm at a distance of
$\sim3.5$~kpc, and was Faraday rotated by the ionized medium along its
path length. Furthermore, they showed that voids in polarization
are due to foreground \ion{H}{2} regions with magnetic fields
disordered on scales of $\sim0.2$~pc, and found a depolarization halo
around the \ion{H}{2} region RCW~94.

Using the RMs from the SGPS test region, \citet{hgm04b} concluded that
a dominant source of fluctuations with an outer scale of a few parsecs
exists in the spiral arms only, and suggested the structure could be
caused by \ion{H}{2}~regions in the arms. This assertion was confirmed
by statistical analysis of RMs of SGPS extragalactic point sources in
the whole SGPS \citep{hbg06a}, which describes the detection of
fluctuations with an outer scale smaller than $\sim$~17~pc in the
Carina and Crux spiral arms, but not present in the interarm regions.

RM data from pulsars and polarized extragalactic point sources in the
SGPS, combined with other available point source data, are being
used to construct a model of the large-scale Galactic magnetic field
(Haverkorn et al.\ 2006b, Brown et al., in prep), discussing the
strength of the regular magnetic field and the abundance and locations
of magnetic field reversals.

The SGPS is an excellent source for finding correlations between radio
(de-)polarization and \ion{H}{2} regions or molecular clouds, which
indicate Faraday rotation effects due to photodissociation regions
(PDRs) around \ion{H}{2} regions \citep{gdm01,wr04}.

Finally, SGPS RM maps reveal individual objects that cannot be
detected in any other way, such as the partial shell centered on
$l=278.5^{\circ}$. These objects do not show any correlation with
H$\alpha$ measurements, indicating that the RM structure cannot be due
to enhanced electron density. Therefore, these objects are likely
purely magnetic, and not detectable otherwise.

\section{Summary and conclusions}

The polarized radio continuum part of the Southern Galactic Plane
Survey (SGPS) provides a major new resource for studies of the ionized
interstellar medium and Galactic magnetic field. The SGPS covers
almost one third of the Galactic plane with a Galactic longitude
coverage of $253\degr < l < 358\degr$, at Galactic latitudes
$|b|<1\degr$. The rms sensitivity of the SGPS continuum data is
$\sim$~0.3~mJy/beam increasing to double that number close to the
Galactic Center, the instrumental polarization is below 0.1\% and the
resolution is 100\arcsec. Due to beam depolarization at low
resolutions, maps with this high resolution reveal a wealth of
polarization structure which would be depolarized in low resolution
maps.

The polarized radio continuum data in the SGPS are uniquely suited for
studies of the ionized gas and magnetic field strength and structure
in the inner Galactic plane, taking into account the constraints of
missing large-scale structure. The polarized intensity can be used for
depolarization studies to obtain the scale of fluctuations in the
magneto-ionized medium. Rotation measure data give information about
the electron-density-weighted magnetic field strength along the line
of sight, and can be used for statistical analysis of these
fluctuations, large-scale modeling of the Galactic magnetic field, or
study of individual magnetic objects.
  
\acknowledgments 

The ATCA is part of the Australia Telescope, which is funded by the
Commonwealth of Australia for operation as a National Facility managed
by CSIRO. This research was supported by the National Science
Foundation through grants AST-0307358 to Harvard College Observatory
and AST-9732695 and AST-0307603 to the University of
Minnesota. M.H. acknowledges support from the National Radio Astronomy
Observatory (NRAO), which is operated by Associated Universities Inc.,
under cooperative agreement with the National Science
Foundation. B.M.G. is supported by an Alfred P. Sloan Research
Fellowship. We would like to thank J. Gelfand for help with observing,
and the staff of the Australia Telescope National Facility for their
support of this project, especially R. Haynes, D. McConnell, R. Sault,
R. Wark, and M. Wieringa.

{\it Facilities:} \facility{ATCA}.


\begin{thebibliography}{}
\bibitem[Broten et al.(1988)]{bmv88}
  Broten, N. W., MacLeod, J. M., \& Vallee, J. P. 1988, Ap\&SS, 141, 303
\bibitem[Brown et al.(2003)]{btj03} 
  Brown, J. C., Taylor, A.~R., \& Jackel, B. J. 2003, ApJS, 145, 213
\bibitem[de Oliveira-Costa et al.(2003)]{dto03}
  de Oliveira-Costa, A., Tegmark, M., O'Dell, C., Keating, B., Timbie,
  P., Efstathiou, G., \& Smoot, G. 2003, PhRvD, 68, 8
\bibitem[Dolag et al.(2005)]{dve05}
  Dolag, K., Vogt, C., \& Ensslin, T. A. 2005, MNRAS, 358, 726
\bibitem[Elmegreen \& Scalo(2004)]{es04}
  Elmegreen, B.~G., \& Scalo, J. 2004, ARA\&A, 42, 211
\bibitem[Ferri\`ere(2001)]{f01} 
  Ferri\`ere, K. M. 2001, RvMP, 73, 1031
\bibitem[Gaensler et al.(2001)]{gdm01} 
  Gaensler, B.~M., Dickey, J.~M., McClure-Griffiths, N.~M., Green,
  A.~J., Wieringa, M.~H., Haynes, R.~F. 2001, ApJ, 549, 959
\bibitem[Hamaker \& Bregman(1996)]{hb96}
  Hamaker, J. P., \& Bregman, J. D. 1996, A\&AS, 117, 161
\bibitem[Haverkorn et al.(2003a)]{hkb03a} 
  Haverkorn, M., Katgert, P., de Bruyn, A. G. 2003a, A\&A, 403, 1031
\bibitem[Haverkorn et al.(2003b)]{hkb03b} 
  Haverkorn, M., Katgert, P., \& de Bruyn, A. G. 2003b, A\&A, 404, 233 
\bibitem[Haverkorn et al.(2004a)]{hkb04a} 
  Haverkorn, M., Katgert, P., \& de Bruyn, A. G. 2004a, A\&A, 427, 549
\bibitem[Haverkorn et al.(2004b)]{hgm04b}
  Haverkorn, M., Gaensler, B.~M., McClure-Griffiths, N.~M., Dickey,
  J.~M., \& Green, A. J. 2004b, ApJ, 609, 776 
\bibitem[Haverkorn et al.(2006a)]{hbg06a} 
  Haverkorn, M., Brown, J. C., Gaensler, B.~M., McClure-Griffiths,
  N.~M., Dickey, J.~M., \& Green, A. J. 2006, ApJL, 637, 33
\bibitem[Haverkorn et al.(2006b)]{hgb06b} 
  Haverkorn, M., Gaensler, B.~M., Brown, J. C., McClure-Griffiths,
  N.~M., Dickey, J.~M., \& Green, A. J. 2006, Ast.~Nach., in press
\bibitem[McClure-Griffith et al.(2005)]{mdg05}
  McClure-Griffiths, N.~M., Dickey, J.~M., Gaensler, B.~M., Green,
  A.~J., Haverkorn, M., Strasser, S. 2005, ApJS, 158, 178 (Paper~I)
\bibitem[McClure-Griffiths et al.(2001)]{mgd01} 
  McClure-Griffiths, N.~M., Green, A.~J., Dickey, J.~M., Gaensler,
  B.~M., Haynes, R.~F., \& Wieringa, M.~H 2001, ApJ, 551, 394 
\bibitem[Morales \& Hewitt(2004)]{mh04}
   Morales, M. F., \& Hewitt, J. 2004 ApJ, 615, 7
\bibitem[Narayan \& Nityananda(1986)]{nn86}
  Narayan, R., \& Nityananda, R. 1986, ARA\&A, 24, 127
\bibitem[Reid(2004)]{r04}
  Reid, R. 2004, in The Magnetized Interstellar Medium, ed.\
  B. Uyan\i ker, W. Reich, R. Wielebinski (Katlenburg-Lindau: Copernicus
  GmbH), 39
\bibitem[Reynolds(1994)]{r94}
  Reynolds, J. E. 1994, ATNF Technical Document Series, Tech. Rep.
  AT/39.3/0400 (Sydney: Australia Telescope National Facility)
\bibitem[Sault et al.(1999)]{sbd99}
  Sault, R. J., Bock, D. C.-J., \& Duncan, A. R. 1999, A\&AS, 139, 387
\bibitem[Sault \& Killeen(2003)]{sk03} 
  Sault, R. J., \& Killeen, N.~E.~B. 2003, The Miriad User's Guide
  (Sydney: Australia Telescope National Facility)
\bibitem[Sault et al.(1996)]{ssb96} 
  Sault, R. J., Staveley-Smith, L., \& Brouw, W. N. 1996, A\&AS, 120,
  375
\bibitem[Scalo \& Elmegreen(2004)]{se04}
  Scalo, J., \& Elmegreen, B.~G. 2004, ARA\&A, 42, 275
\bibitem[Simard-Normandin et al.(1981)]{skb81} 
  Simard-Normandin, M., Kronberg, P. P., \& Button, S. 1981, ApJS, 45,
  97
\bibitem[Stanimirovi\'c (2002)]{s02} 
  Stanimirovi\'c, S. 2002, in ASP Conference Proceedings, Vol. 278,
  eds.\ S. Stanimirovi\'c, D. Altschuler, P. Goldsmith, and C. Salter,
  (San Francisco: Astronomical Society of the Pacific) p. 375-396
\bibitem[Steer et al.(1984)]{sdi84}
  Steer, D. G., Dewdney, P. E., \& Ito, M. R. 1984, A\&A, 137, 159
\bibitem[Taylor et al.(2003)]{tgp03} 
  Taylor, A.~R., Gibson, S.~J., Peracaula M., Martin, P.~G.,
  Landecker, T.~L., Brunt, C.~M., Dewdney, P.~E., Dougherty, S.~M.,
  Gray, A.~D., Higgs, L.~A., Kerton, C.~R., Knee, L.~B.~G., Kothes,
  R., Purton, C.~R., Uyan\i ker, B., Wallace, B.~J., Willis, A.~G., \&
  Durand, D. 2003, AJ, 125, 3145
\bibitem[Testori et al.(2004)]{trr04}
  Testori, J. C., Reich, P., \& Reich, W. 2004, in The Magnetized
  Interstellar Medium, ed.\ B. Uyan\i ker, W. Reich, R. Wielebinski
  (Katlenburg-Lindau: Copernicus GmbH), 57
\bibitem[Uyan\i ker et al.(1998)]{ufr98}
  Uyan\i ker, B., F\"urst, E., Reich, W., Reich, P., \& Wielebinski,
  R. 1998, A\&AS, 132, 401	
\bibitem[Uyan\i ker et al.(2003)]{ulg03}
  Uyan\i ker, B., Landecker, T. L., Gray, A. D., \& Kothes, R. 2003,
  ApJ, 585, 785
\bibitem[Vogt et al.(2005)]{vde05}
  Vogt, C., Dolag, K., \& Ensslin, T. A. 2005, MNRAS, 358, 732
\bibitem[Wardle \& Kronberg(1974)]{wk74}
  Wardle, J. F. C., \& Kronberg, P. P. 1974, ApJ, 194, 249
\bibitem[Wolleben et al.(2005)]{wlr05}
  Wolleben, M., Landecker, T. L., Reich, W., \& Wielebinski, R., 2005,
  A\&A, in press, astro-ph/0510456
\bibitem[Wolleben \& Reich(2004)]{wr04}
  Wolleben, M., \& Reich, W. 2004, A\&A, 427, 537
\end{thebibliography}
\end{document}